\documentclass[a4paper]{jpconf}
\usepackage{graphicx,epsfig}
\usepackage{subeqn}
\usepackage[usenames]{color}
\usepackage{ulem} 
\usepackage{fancyhdr}

\newcommand{\eF}{{\epsilon_{\rm F}}}
\newcommand{\kB}{{k_{\rm B}}}

\newcommand{\cN}{{\mathcal N}}
\newcommand{\cZ}{{\mathcal Z}}

\begin{document}
\title{Universal features in the thermodynamics and heat transport by particles of any statistics}

\author{Drago\c s-Victor Anghel}

\address{Department of Theoretical Physics, National Institute of Physics and Nuclear Engineering, 30 Reactorului Street, P.O.BOX MG-6, RO-077125 Magurele, Jud. Ilfov, ROMANIA}

\ead{dragos@theory.nipne.ro}

\begin{abstract}
I discuss in parallel two universal phenomena: the independence of statistics of the heat capacity and entropy of ideal gases of the same, constant, density of states, on one hand, and the independence of statistics of the heat and entropy  transport through one-dimensional channels, on the other hand. I show that there is a close similarity between the microscopic explanations of each of these phenomena.
\end{abstract}

\section{Introduction}

The \textit{thermodynamic equivalence} is a name coined by M. H. Lee \cite{PhysRevE.55.1518.1997.Lee} for the property of ideal Bose and Fermi gases confined in two-dimensional (2D) boxes to have the same heat capacity, $C_V$, and entropy, $S$. This implies that from the thermodynamic perspective, 2D Bose and Fermi gases under canonical conditions (i.e. has a constant volume and can exchange only heat with the environment) are identical.

This property was first proved by Auluck and Kothari in Ref. \cite{ProcCambrPhilos42.272.1946.Auluc} and then rediscovered by May \cite{PhysRev.135.A1515.1964.May} and Viefers, Ravndal and Haugset \cite{AmJPhys63.369.Viefers}. Nevertheless, it did not receive much attention until  Lee introduced a unitary description of ideal Bose and Fermi gases \cite{PhysRevE.55.1518.1997.Lee,JMathPhys36.1217.1995.Lee,PhysA304.421.2002.Lee,ActaPhysicaPolonicaB.40.1279.2009.Lee} in terms of polylogarithmic functions \cite{Lewin:book,PhysRevE.56.3909.1997.Lee}.

In Ref. \cite{PhysRevLett.67.937.1991.Haldane}, Haldane introduced an extension of the Pauli exclusion principle, which was latter called the \textit{fractional exclusion statistics} (FES). In FES, the system is divided into species of particles, that we shall identify by an index, e.g. $i=0,1,\ldots$. In each species there are $G_i$ available single-particle states and $N_i$ particles. The FES is defined by the fact that if we change for example $N_i$ by $\delta N_i$, then the dimension of species $j$, for any $j=0,1,\ldots$, changes by $\delta G_j=-\alpha_{ji}\delta N_i$, where $\alpha_{ij}$ are called the FES parameters. The diagonal parameters ($\alpha_{ii}$) are called \textit{direct parameters}, whereas the rest of them ($\alpha_{ij},\ i\ne j$) are called the \textit{mutual parameters}. For example for a continuous (i.e. macroscopic) FES system the species of particles are defined by coarse-graining the phase space, each grain representing a species of particles \cite{JPhysA.40.F1013.2007.Anghel,PhysLettA.372.5745.2008.Anghel,RJP.54.281.2009.Anghel}. 

The general properties of the FES parameters were deduced only relatively recently \cite{EPL.87.60009.2009.Anghel,PhysRevLett.104.198901.2010.Anghel,PhysRevLett.104.198902.2010.Wu}. A quite general ansatz for the FES parameters that obey the properties introduced in \cite{EPL.87.60009.2009.Anghel} is \cite{JPhysA.40.F1013.2007.Anghel,EPL.90.10006.2010.Anghel}
\begin{equation}
\alpha_{ij}=\alpha^{(e)}_{ij}+\alpha^{(s)}_{i}\delta_{ij}.
\label{alpha_nd}
\end{equation}
The parameters $\alpha^{(e)}_{ij}$ are ``extensive'', i.e. they are proportional to $G_i$ -- the dimension of the species on which they act,
\begin{equation}
\alpha^{(e)}_{ij}\equiv a_{ij}G_i, \label{def_alpha_e}
\end{equation}
whereas the parameters $\alpha^{(s)}_{i}$ are direct parameters and are not extensive. 

Typically, in the literature we find exclusion statistics parameters of the $(s)$ type (see e.g. 
\cite{PhysRevLett.73.922.1994.Wu,PhysRevLett.73.3331.1994.Murthy,PhysRevLett.74.3912.1995.Sen,PhysRevB.60.6517.1999.Murthy,JPhysB33.3895.2000.Bhaduri,PhysRevLett.86.2930.2001.Hansson,JPA35.7255.2002.Anghel,RomRepPhys59.235.2007.Anghel}), so in general $\alpha_{ij}=0$ for any $i\ne j$. If, moreover, we impose $\alpha_{ii}\equiv\alpha$ for any $i$, the thermodynamic calculations simplify considerably. If $\alpha=0$ we obtain the Bose statistics, whereas if $\alpha=1$ we obtain the Fermi statistics. 

If $\alpha_{ij}\equiv\alpha\delta_{ij}$, one obtains a very surprising result: in 2D systems $C_V$ and $S$ are independent of $\alpha$ and equal to the heat capacity and entropy of ideal Bose and Fermi gases. Therefore 2D systems of \textit{any} statistics, i.e. Bose, Fermi or FES, are thermodynamically equivalent \cite{JPA35.7255.2002.Anghel}.

The property which leads to the thermodynamic equivalence of 2D ideal gases is the constant density of single-particle states (DOS). A system of ideal particles in a $d$-dimensional ($d$D) box has a DOS of the form 
\begin{equation} \label{sigma_d}
\sigma_d(\epsilon)=V\frac{d}{(2\pi)^{d/2}\Gamma\left(\frac{d}{2}+1 \right)}\left(\frac{m}{\hbar^2}\right)^{d/2} \epsilon^{(d/2)-1} \equiv C \epsilon^{(d/2)-1} ,
\end{equation}
where $\epsilon$ is the single-particle energy, $V$ si the $d$D volume, $m$ is the mass of the particle and $\Gamma$ is the gamma function. If $d=2$, then $\sigma_{d=2}(\epsilon)=Vm/(\pi\hbar^2)\equiv C$ is a constant. We shall see in section~\ref{thermodynamics} that the ideal systems of any statistics, of the same, constant DOS are thermodynamically equivalent.

The thermodynamic equivalence has a correspondent in the transport properties of ideal gases. At first, Rego and Kirczenow observed that in the low temperature limit, the heat conductivity of ideal particles confined in a 1D channel is independent of the dispersion relation  \cite{PhysRevLett.81.232.1998.Rego} and, moreover, is independent also of their statistics  \cite{PhysRevB.59.13080.1999.Rego}. The statistics independence of the heat conductivity implies the independence of statistics of the entropy current in 1D channels, as it was shown by Blencowe \cite{PhysRep.395.159.2004.Blencowe}. 

The independence of the low temperature heat conductivity of the dispersion relation have been confirmed experimentally in 1D phonons \cite{Nature404.974.2000.Schwab} and photons \cite{Nature.444.187.2006.Meschke,PhysRevLett.102.200801.2009.Timofeev} channels. 

The low temperature results of Rego and Kirczenow have been extended in Ref. \cite{EPL.94.60004.2011.Anghel}. In this paper we pointed to another class of universality in systems of ideal particles, namely we showed that the low temperature statistics independence of the 1D channel heat conductivity can be extended to any temperature. In other words, the heat conductivity of ideal particles through a 1D channel is independent of the statistics of the particles at any temperature. 

In principle all the universal physical properties should have also an universal (and hopefully very simple) microscopic physical explanation; there should be a key, basic feature, common to all these systems, that is leading always to similar results. For the thermodynamic equivalence the key feature is the spectrum of excitations, which is independent of statistics \cite{JPA35.7255.2002.Anghel}, whereas for the 1D universality the key feature is the current of excitations, which is also independent of statistics \cite{EPL.94.60004.2011.Anghel}. The two microscopic explanations are very closely related and, as we shall see in the next two sections, the quantities related to the 1D transport have mathematical expressions very similar to the ones that describe the equilibrium thermodynamics.

In section \ref{thermodynamics} we shall discuss the thermodynamics of systems of constant DOS and we shall present the main results and formulae. We shall also present the microscopic explanation of the thermodynamic equivalence. 

In section \ref{transport} we shall present the calculations of the heat, particle and entropy transport in 1D channels and, by employing the formulae of section~\ref{thermodynamics}, we shall show that they are independent of statistics at any temperature. The microscopic interpretation of this phenomenon is given in the end of section~\ref{transport} and is very similar to the microscopic interpretation of the thermodynamic equivalence. 

Section \ref{conclusions} is reserved for conclusions.

\section{Thermodynamic equivalence}\label{thermodynamics}

Let us assume that we have a macroscopic system of ideal FES particles, with a DOS of the form $\sigma(\epsilon)=C\epsilon^s$. The FES parameters are $\alpha_{ij}\equiv\alpha\delta_{ij}$, for any $i$ and $j$. We associate the same energy to all the particles in a species ($\epsilon_i$, $i=0,1,\ldots$) and the same chemical potential, $\mu$, to all the species. If we denote by $G_{i}$ the number of states in the species $i$ when there are no particles in it, then in the presence of $N_i$ particles the number of available states becomes $G_{i}-\alpha N_i$ \cite{PhysRevLett.67.937.1991.Haldane,PhysRevLett.73.922.1994.Wu,PhysRevB.60.6517.1999.Murthy}. The number of particle arrangements that we can have in the species $i$ is
\begin{equation}
  W(G_{i},N_i) = \frac{\left[G_i+(1-\alpha)N_i-1\right]!} {N_i!\left[G_i-\alpha N_i-1\right]!}. \label{WGiNi}
\end{equation}
Therefore the total number of microscopic configurations in the system, when the population of each species is fixed, is given by the product \cite{PhysRevLett.73.922.1994.Wu}
\begin{equation}
W(\{G_{i},N_i\}) = \prod_i \frac{\left[G_i+(1-\alpha)N_i-1\right]!} {N_i!\left[G_i-\alpha N_i-1\right]!}. \label{Wtot}
\end{equation}

The grandcanonical partition function of the system, at temperature $T$ and chemical potential $\mu$, is 
\begin{equation}
\cZ = \sum_{\{N_i\}}\cZ(\{G_{i},N_i\}) = \sum_{\{N_i\}}W(\{G_{i},N_i\})\exp\left[\sum_i \beta N_i(\mu-\epsilon_i) \right], \label{cZNi}
\end{equation}
where $\beta=1/(\kB T)$ and $\cZ(\{G_i,N_i\})$ is the ``partition function'' corresponding to a particular choice of the set $\{N_i\}$.

The standard procedure to find the particle populations of the species, denoted by $n_i\equiv N_i/G_{i}$, is to maximize $\cZ(\{G_{i},N_i\})\equiv\cZ(\{G_{i},n_i\})$ with respect to the $n_i$'s: $\partial \cZ(\{G_{i},n_i\})/ \partial n_i =0$. In this way it was obtained \cite{PhysRevLett.73.922.1994.Wu}
\begin{subequations} \label{syst_nw_alpha}
\begin{equation} \label{n}
n(\epsilon) = \{w(\zeta_\epsilon) + \alpha \}^{-1} , 
\end{equation}
where $w$ is a function that satisfies the equation
\begin{equation}\label{w}
  w(\zeta_\epsilon)^\alpha[1+w(\zeta_\epsilon)]^{1-\alpha}=\zeta_\epsilon^{-1} \equiv \rme^{\beta(\epsilon - \mu)}
\end{equation}
\end{subequations}

Now we apply the results above to our system of DOS $\sigma(\epsilon)=C\epsilon^s$. From (\ref{cZNi}) we write the expression for the grandcanonical thermodynamic potential, 
\begin{subequations} \label{eqs_Omega}
\begin{equation}
  -PV = \Omega=-\kB T\log(\cZ) = k_{\rm B}T \int_0^\infty \rmd\epsilon\, C\epsilon^s \log{\{(1-\alpha n)/[1+(1-\alpha)n]\}}. \label{Omega0}
\end{equation}
Integrating by parts and using the equations (\ref{syst_nw_alpha}) we obtain 
\begin{equation}\label{pv}
PV = \frac{1}{s+1}\int_0^\infty \rmd\epsilon\, C\epsilon^{s+1} n(\epsilon) 
\equiv \frac{U}{s+1} \, ,
\end{equation}
\end{subequations}
where $U$ is the internal energy of the system. 

Similarly, the total number of particles is 
\begin{equation}
  N = \int_0^\infty \rmd\epsilon\, C\epsilon^s n(\epsilon). \label{Eq_N}
\end{equation}

All these functions may be calculated for example by expressing $\epsilon$ in terms of $w$,
\begin{equation}
  \beta\epsilon = \log{[w^\alpha (1+w)^{1-\alpha}]} + \beta\mu, \label{eps_w}
\end{equation}
but the integrals cannot be performed analytically for general $s$ and $T$. 

For $s=0$ ($\sigma\equiv C$), all the thermodynamic quantities can be expressed in terms of elementary or polylogarithmic functions \cite{JPA35.7255.2002.Anghel}. I start with $N$, which, by simple algebra may be written as
\begin{equation} \label{N}
N = k_{\rm B}T \sigma \log{(1+y_0)} ,
\end{equation}
where $y_0\equiv 1/w(\zeta_{\epsilon=0})$ satisfies the equation 
\begin{equation} \label{Eq_y0}
  (1+y_0)^{1-\alpha}/y_0 = \zeta^{-1}_{\epsilon=0} \equiv \rme^{-\beta\mu}.
\end{equation}
%

\begin{figure}[t]
  \begin{center}
    \includegraphics[width=60mm]{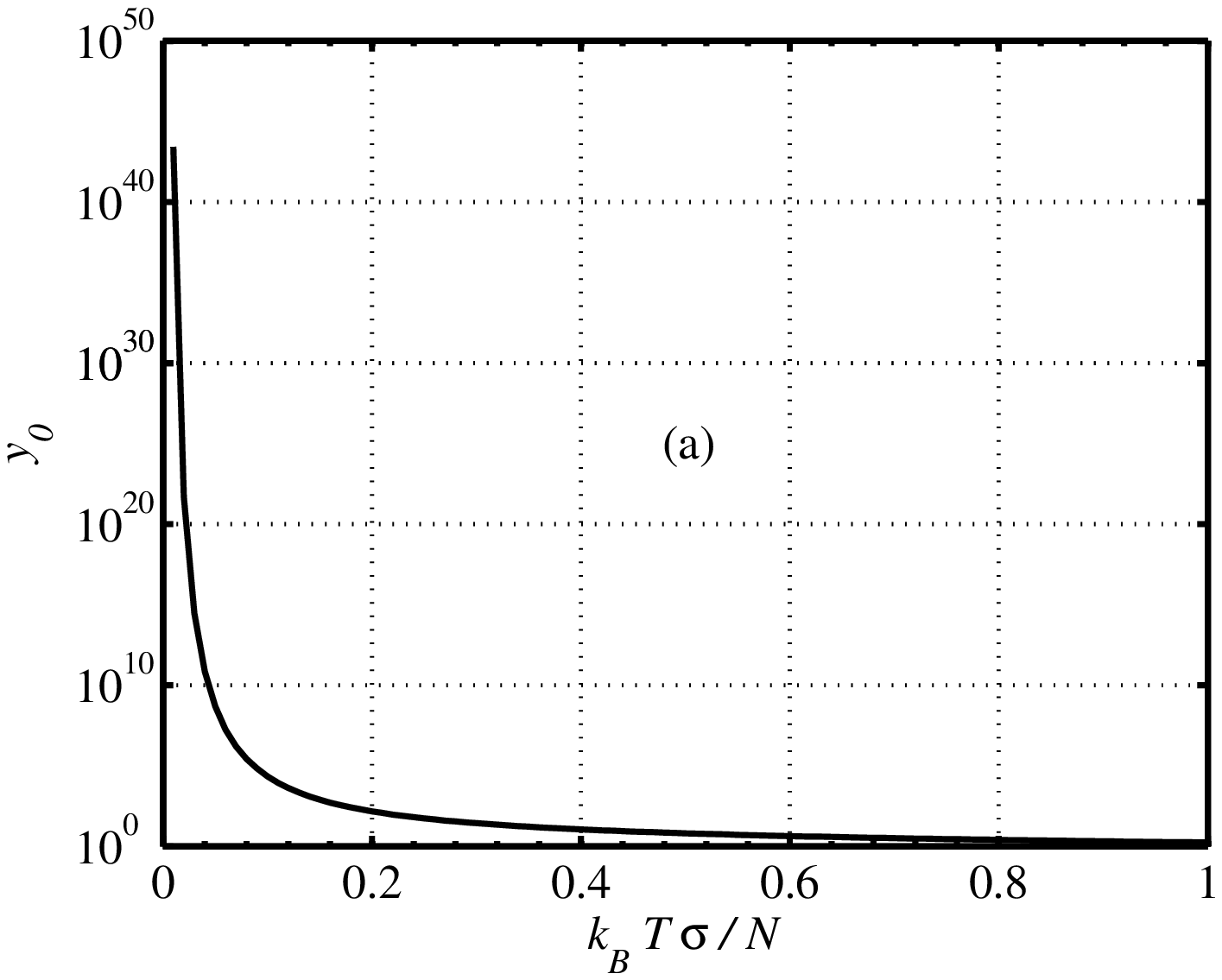} \includegraphics[width=60mm]{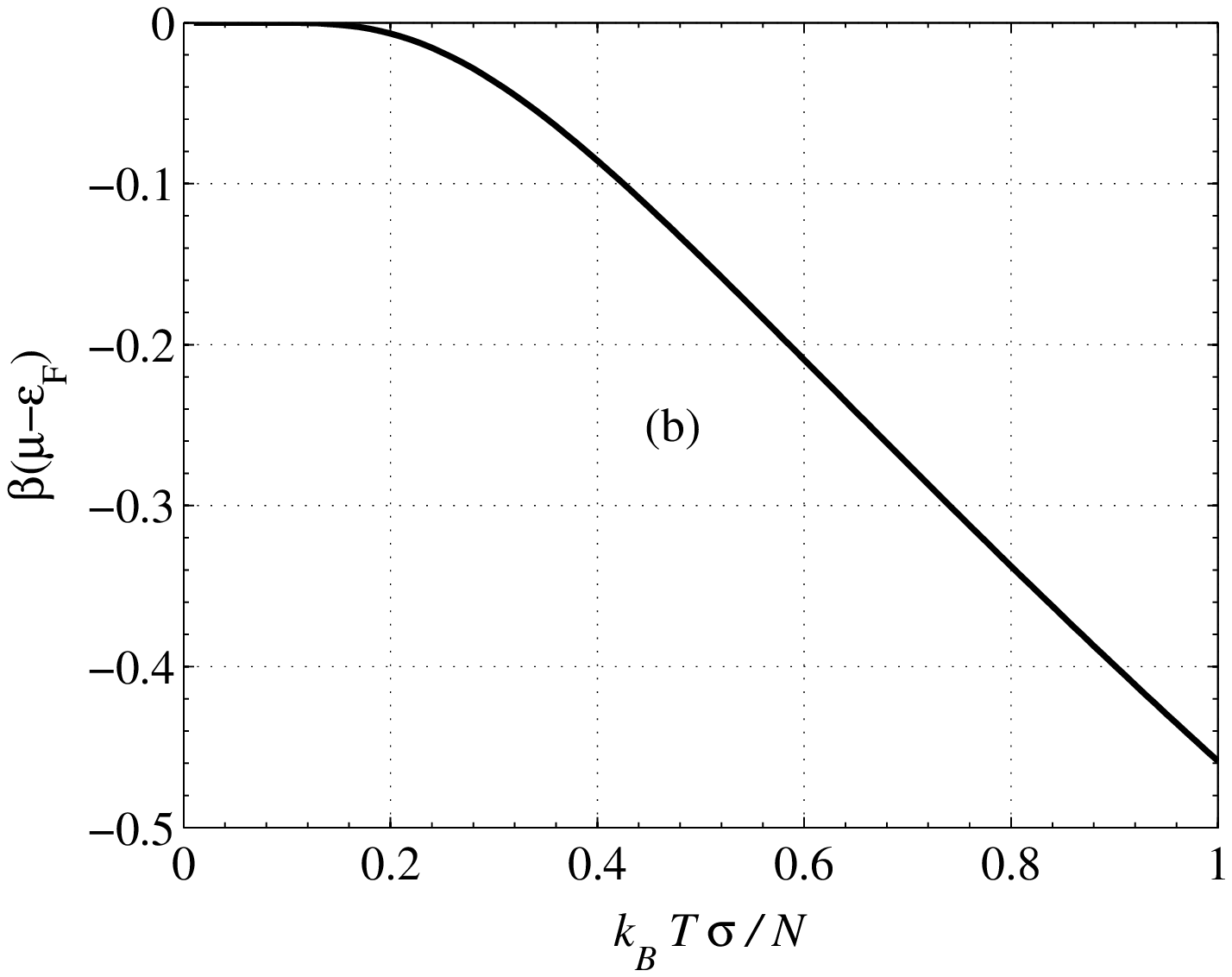}
  \end{center}
  \caption{The statistics independent functions, $y_0$ (a) and $\beta(\mu-\eF)$ (b), vs the scaled temperature, $\kB T\sigma/N$.}    
  \label{fig_y0_mu_vs_T}
\end{figure}

We observe from eq. (\ref{N}) that $y_0$ is a function of the dimensionless variable $x\equiv\kB T\sigma/N$, and does not depend on $\alpha$. Moreover, as seen in figure \ref{fig_y0_mu_vs_T}a, $y_0(x)$ is a monotonically decreasing function, with $y_0(x\to\infty)=0$ and $y_0(x\to0)\to\infty$. 

From the equations (\ref{N}) and (\ref{Eq_y0}) follows
\begin{equation}\label{mu}
\exp{[(\mu - \alpha N/\sigma)/k_{\rm B} T]} = 1- \exp{[-N/(\sigma k_{\rm B}T)]},
\end{equation}
where we can identify the (generalized) Fermi energy as $\epsilon_{\rm F} \equiv \lim_{T\to 0} \mu = \alpha N/\sigma$ and observe that $\mu - \epsilon_{\rm F}$ is also independent of $\alpha$ (see figure \ref{fig_y0_mu_vs_T}b).

After some algebra we obtain 
\begin{eqnarray} 
-\Omega = U &=& -(k_{\rm B}T)^2 \sigma\left[\frac{1-\alpha}{2} \log^2{(1+y_0)} + Li_2(-y_0)\right] \nonumber \\ 
&=& (k_{\rm B}T)^2 \sigma\left[\frac{\alpha}{2} \log^2{(1+y_0)} + Li_2\left(\frac{y_0}{1+y_0}\right)\right] \label{omega1}
\end{eqnarray}
where $Li_2(x)=\sum_{k=1}^\infty x^k/k^2$ is Euler's dilogarithm \cite{Lewin:book}. The two expressions for $U$ in (\ref{omega1}) are connected by Landen's relation, $Li_2(x)+Li_2[-x/(1-x)] = -(1/2)\log^2{(1-x)}$, which is valid for any $x<1$ \cite{Lewin:book}. 


Using (\ref{N}) into (\ref{omega1}) we obtain
\begin{subequations}\label{echiv}
\begin{eqnarray} 
-\frac{\sigma}{N^2}\Omega = \frac{\sigma}{N^2}U &=& - \left(\frac{k_{\rm B} T\sigma}{N}\right)^2 Li_2(-y_0) + \frac{\alpha}{2} - \frac{1}{2} \label{echivF} \\
&=& \left(\frac{k_{\rm B} T\sigma}{N}\right)^2 Li_2\left(\frac{y_0}{1+y_0}\right) + \frac{\alpha}{2} \label{echivB}
\end{eqnarray} 
\end{subequations}
In the expression (\ref{echivF}), $U_{\rm F} = -(k_{\rm B} T)^2 \sigma Li_2(-y_0)$ is the internal energy of a Fermi gas ($U=U_{\rm F}$, if $\alpha=1$), whereas in the expression (\ref{echivB}), $U_{\rm B} = (k_{\rm B} T)^2 \sigma Li_2[y_0/(1+y_0)]$, represents the internal energy of a Bose gas ($U=U_{\rm B}$, if $\alpha=0$).  

Since $y_0$ does not depend on $\alpha$, but just on $\kB T\sigma/N$ and $T$, it is obvious that the differences between the thermodynamic potentials of gases with different $\alpha$'s come just from an additive constant. All the temperature dependence is the same. 

To calculate $C_V$, we use the expression 
\begin{equation}
 C_V = \frac{\partial U}{\partial T} - \frac{\partial U}{\partial \mu}\frac{\partial N}{\partial T}\left(\frac{\partial N}{\partial \mu}\right)^{-1} \label{expr_gen_CV}
\end{equation}
and obtain 
\begin{subequations} \label{equivCVS}
\begin{equation}
 C_{\rm V} = - \frac{N^2}{T\sigma}\frac{1+y_0}{y_0}- 2k_{\rm B}^2 
T\sigma Li_2(-y_0) . \label{cv} 
\end{equation}

To calculate the entropy we use 
%
  $S=U/T+PV/T-\mu N/T$, 
%
together with equations (\ref{N}), (\ref{Eq_y0}) and (\ref{echiv}), to obtain
\begin{eqnarray}
S &=& -k_{\rm B}^2 T \sigma [2Li_2(-y_0) + \log{(1+y_0)}\log{y_0}]. \label{entropia}
\end{eqnarray}
\end{subequations}
Equations (\ref{cv}) and (\ref{entropia}) are independent of $\alpha$ and they express eventually in the most compact way the thermodynamic equivalence. 

\begin{figure}[t]
  \begin{center}
    \includegraphics[width=70mm]{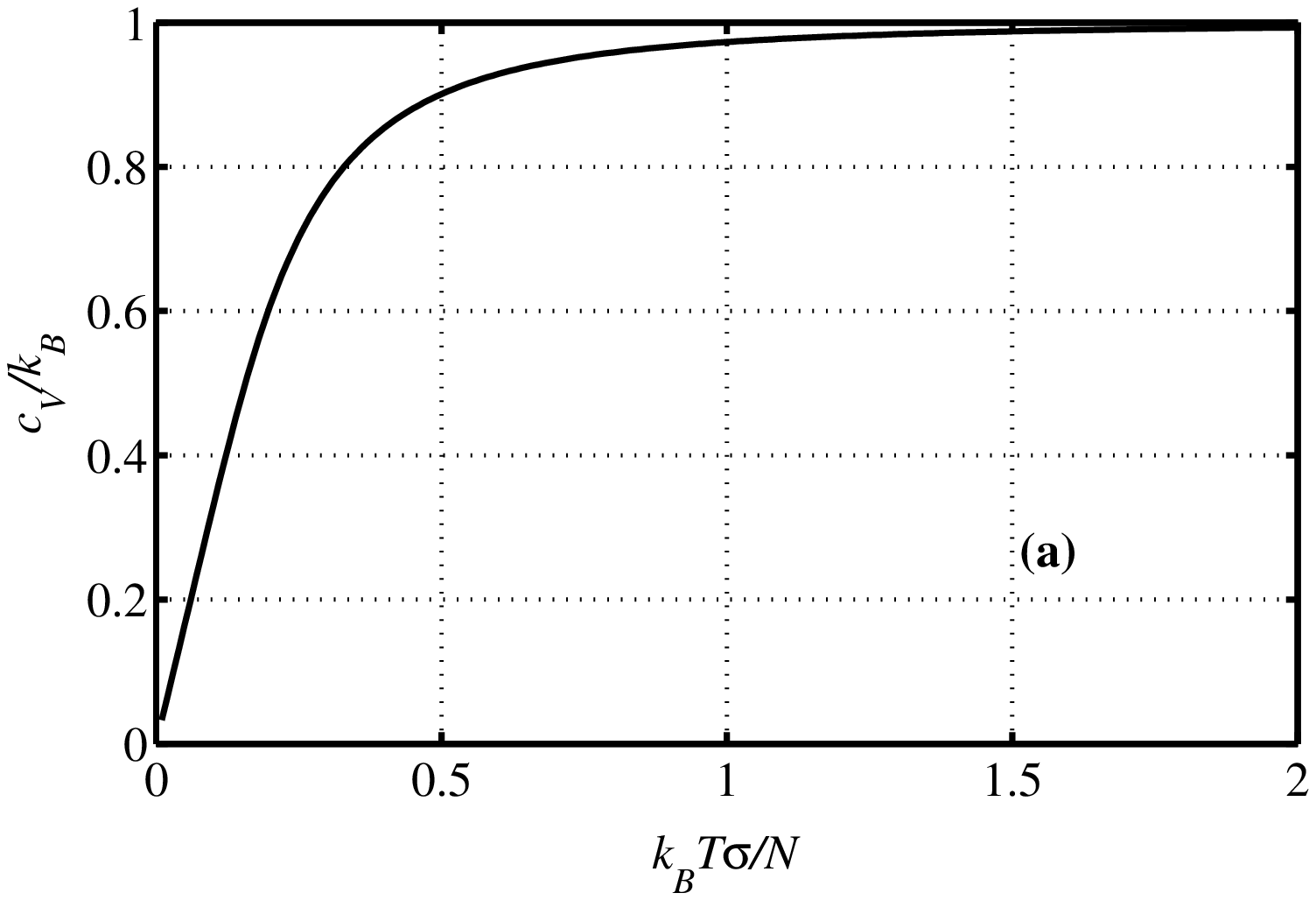} \includegraphics[width=70mm]{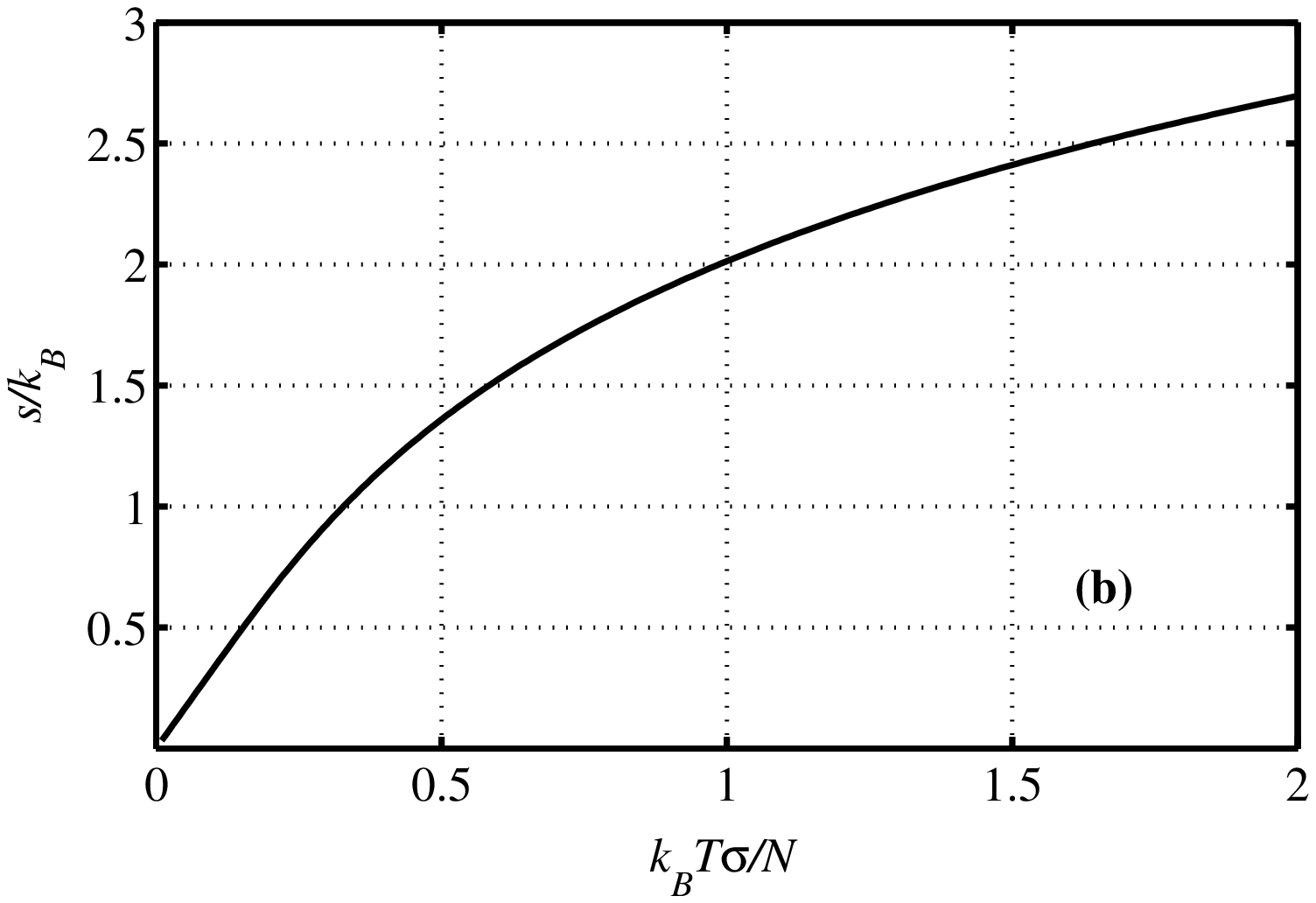}
  \end{center}
  \caption{The specific heat (a), $c_V/\kB\equiv C_V/(N\kB)$, and entropy per particle (b), $s/\kB\equiv S/(N\kB)$, for a system of particles of constant DOS, $\sigma$, and any $\alpha$.}
  \label{cV_S_vs_T}
\end{figure}

In figure \ref{cV_S_vs_T} (a) and (b) we plotted $c_V/\kB\equiv C_V/(N\kB)$ and $s/\kB\equiv S/(N\kB)$, respectively. 

Another way to calculate the entropy is by using the definition 
\begin{equation}
 TS = \kB T\log W , \label{entropy_conf}
\end{equation}
where $W$ is given by (\ref{Wtot}). From the expression (\ref{Wtot}) we take only the most probable distribution, i.e. the one that satisfies the equations (\ref{syst_nw_alpha}), and, after some algebra we obtain 
\begin{eqnarray}
 \log W &=& \sum_i G_i\left\{[1+(1-\alpha)n_i]\log[1+(1-\alpha)n_i] - n_i\log n_i - [1-\alpha n_i]\log[1-\alpha n_i] \right\} \nonumber \\
  &=& \sigma\int_0^\infty\rmd\epsilon\{[1+(1-\alpha)n(\epsilon)]\log[1+(1-\alpha)n(\epsilon)] - n(\epsilon)\log n(\epsilon) \nonumber \\
  && - [1-\alpha n(\epsilon)]\log[1-\alpha n(\epsilon)]\} \label{Wtot_int}
\end{eqnarray}
where we have used the Stirling approximation, $\log(N!)\approx N\log(N/\rme)$, valid for $N\gg1$, and neglected the terms of the order of $1/G_i$. Plugging (\ref{syst_nw_alpha}) into (\ref{Wtot_int}) we obtain
\begin{eqnarray}
 TS &=& \kB T\sigma\int_0^\infty\frac{(w+1)\log(w+1)-w\log w}{w+\alpha}\rmd\epsilon \nonumber \\
  &=& (\kB T)^2\sigma\int_{y_0^{-1}}^\infty\left[\frac{\log(w+1)}{w}-\frac{\log w}{w+1}\right]\rmd\epsilon \nonumber \\
  &=& (\kB T)^2\sigma\int_0^{y_0}\left[\frac{\log(y+1)}{y}-\frac{\log y}{y+1}\right]\rmd\epsilon \nonumber \\
  &=& -(\kB T)^2\sigma[2Li_2(-y_0) + \log{(1+y_0)}\log{y_0}] , \label{SWtot_int}
\end{eqnarray}
which is equation (\ref{entropia}), as expected. 

Since according to Eq. (\ref{N}) $y_0\to \infty$ when $T\to 0$, making use of the asymptotic behaviour of the dilogarithm, $Li_2(-y_0) \sim -[\pi^2/6 + \log^2(y_0)/2]$ one can show that in the limit of low temperatures 
\begin{equation}
  C_{\rm V} \sim (\pi^2/3)k_{\rm B}^2T\sigma\{1- O[\log^2(1+y_0)/y_0]\}. \label{cv_limT0}
\end{equation}

\subsection{Microscopic interpretation} \label{thermo_micro}

The microscopic interpretation of the remarkable identities expressed by Eqs. (\ref{equivCVS}) was given in Ref. \cite{JPA35.7255.2002.Anghel}. We showed in that paper that if we have two FES systems of the same, constant, DOS, but different $\alpha$'s, one can establish a one-to-one correspondence between configurations of particles in the two systems that have the same excitation energy. The excitation energy is the difference between the energy of the system in the given configuration and the energy of the ground state; the average of the excitation energy is $U_{\rm B}$, defined above.

\begin{figure}[t]
  \begin{center}
    \includegraphics[width=80mm]{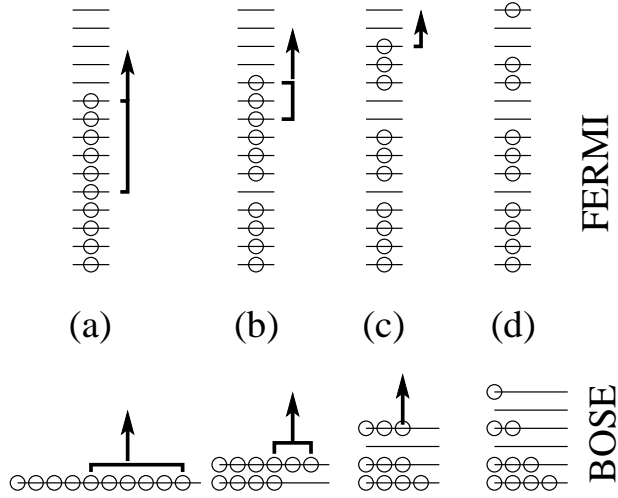}
  \end{center}
  \caption{Excitations in a Bose and a Fermi gas, each of 10 particles and the same DOS. To each configuration of bosons it corresponds a configuration of fermions, with the same excitation energy. In (a) the two systems are in the ground state; in (b) six particles in each are excited on the first ``level''; from the first level, 3 particles are excited two levels up (c), whereas in (d), one particle from the uppermost ``level'' is excited two states up.}    
  \label{excitations2011_fig}
\end{figure}

In figure \ref{excitations2011_fig} we give an example of correspondence between configurations of fermions and configurations of bosons with the same excitation energies. In this figure we can see how the correspondence is established. Based on this method, in Ref. \cite{RomJPhys.53.689.2008.Anghel} we showed how we can transform a general system of fermions into a system of bosons. 

\section{Transport equivalence}\label{transport}

In Ref. \cite{PhysRevB.59.13080.1999.Rego}, Rego and Kirczenow proved theoretically that the low temperature limit of the 1D heat conductance if ideal particles is equal to $\kappa_0\equiv\pi^2\kB ^2T/(3h)$ and is independent of the statistics of the particles. We extended this result in Ref. \cite{EPL.94.60004.2011.Anghel} and showed that the heat conductivity of 1D channels is independent of statistics at any temperature.

\begin{figure}[t]
  \begin{center}
    \includegraphics[width=70mm]{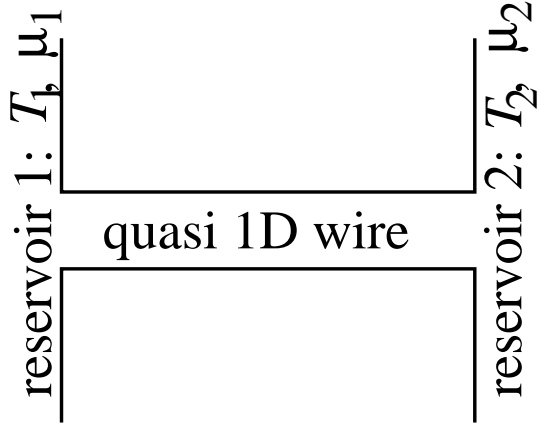}
  \end{center}
  \caption{Schematic view of the heat and particle transport experiment. The system, which consists of two reservoirs connected by a quasi-1D wire, contains FES particles of parameter $\alpha$. The two reservoirs are at the temperatures and chemical potentials $(T_1,\mu_1)$ and $(T_2,\mu_2)$, respectively.}
  \label{q1D_channel_scheme}
\end{figure}

Let us now consider a two-terminal transport experiment in which the reservoirs 1 and 2 are connected by a quasi 1D wire, like in figure \ref{q1D_channel_scheme}. The temperatures and chemical potentials in the two reservoirs will be denoted by $T_{r}$ and $\mu_{r}$, respectively ($r=1,2$). The particles in the system obey FES of parameter $\alpha$. If there are $\cN_c$ transport channels through the wire, the Landauer formula for the particle and energy fluxes between the two reservoirs are \cite{PhysRevB.59.13080.1999.Rego,PhysRep.395.159.2004.Blencowe}
\begin{subequations}\label{IUgen}
\begin{eqnarray}
  I &=& \sum _{p=1}^{\cN_c}\int_{0}^{\infty}\frac{dk}{2\pi}v_{p}(k)\left[n_{1}(k)-n_{2}(k)\right]\zeta_p(k) , \label{Igen} \\
  \dot U &=& \sum _{p=1}^{\cN_c}\int_{0}^{\infty}\frac{dk}{2\pi}\epsilon v_{p}(k)\left[n_{1}(k)-n_{2}(k)\right]\zeta_p(k) , \label{Ugen}
\end{eqnarray}
\end{subequations}
respectively, where $v_{p}(k)$ is the group velocity of the particles of momentum $k$ and $n_{r}(k)$ denotes the thermal particle population in the reservoir $r$, i.e. $n_r(k)\equiv n[\epsilon_r(k),\mu_r,T_r]$ and satisfies Eqs. (\ref{syst_nw_alpha}); finally, $\zeta_p(k)$ is the particle transmission coefficient through the wire.

Following \cite{EPL.94.60004.2011.Anghel} and using the thermodynamic relation for heat exchange, $Q=T\delta S=\delta U-\mu\delta N$ (see in section \ref{thermodynamics} the relation for $S$), where $\delta S$ is the entropy variation in the process, we write the heat flux, 
\begin{equation}
 \dot Q=\dot U-\mu I \label{dotQ1}
\end{equation}
For massless particles, like phonons, $\dot U\equiv\dot Q$, since $\mu=0$. 

Since the particle group velocity is $v_{p}(k)=\hbar^{-1}(d\epsilon(k)/dk)$, and assuming further that $\zeta_p(k)\equiv\zeta_p$ is independent of $k$, Eqs. (\ref{IUgen}) get the simple form 
\begin{subequations}\label{IUgen1}
\begin{eqnarray}
  I &=& \sum _{p=1}^{\cN_c}\frac{\zeta_p}{h}\int_{\epsilon_p(0)}^{\infty}d\epsilon \left[n(\epsilon,\mu_1,T_1)-n(\epsilon,\mu_2,T_2)\right]
  \label{Igen1} \\
  \dot U &=& \sum _{p=1}^{\cN_c}\frac{\zeta_p}{h}\int_{\epsilon_p(0)}^{\infty}d\epsilon\,\epsilon\left[n(\epsilon,\mu_1,T_1)-n(\epsilon,\mu_2,T_2)\right]
  \label{Ugen1}
\end{eqnarray}
\end{subequations}
where $\epsilon_p(0)$ is the lowest energy level in the channel $p$. 

We are working in the linear regime, i.e. $|T_1-T_2|/T_r\ll1$ and $|\mu_1-\mu_2|\ll \kB T_r$  ($r=1,2$), so we shall denote the average temperature, $(T_1+T_2)/2$, by $T$. We introduce the notations $I_1(\mu,T)=\sum_p I_{p,1}(\mu,T)$ and $\dot U_1(\mu,T)=\sum_p\dot U_{p,1}(\mu,T)$, with 
\begin{subequations}\label{IUnsplit}
\begin{eqnarray}
  I_{p,1}(\mu,T)&\equiv&\frac{\zeta_p}{h}\int_{\epsilon_p(0)}^{\infty}n_{\alpha}(\epsilon,\mu,T)d\epsilon , \label{Insplit} \\
  \dot U_{p,1}(\mu,T)&\equiv&\frac{\zeta_p}{h}\int_{\epsilon_p(0)}^{\infty}\epsilon n_{\alpha}(\epsilon,\mu,T)d\epsilon ,
  \label{Unsplit}
\end{eqnarray}
\end{subequations}
and define $I$ and $\dot U$ as 
\begin{subequations}\label{dIU} 
\begin{eqnarray}
  I &\equiv& \delta I_{1} = \frac{\partial I_{1}}{\partial T}\delta T + \frac{\partial I_{1}}{\partial \mu}\delta \mu, \label{dI} \\
  \dot U &\equiv& \delta \dot U_{1} = \frac{\partial \dot U_{1}}{\partial T}\delta T + \frac{\partial \dot U_{1}}{\partial \mu}\delta \mu. \label{dU} 
\end{eqnarray}
\end{subequations}

Rego and Kirczenow calculated the low temperature limit of $\kappa_{\dot U}\equiv\partial \dot U_{1}/\partial T$ -- which we shall call here the energy conductance -- and obtained $\kappa_{\dot U}\stackrel{T\to0}{\sim}\cN_c\kappa_0$ for all $\alpha$'s \cite{PhysRevB.59.13080.1999.Rego,PhysRep.395.159.2004.Blencowe}, if $\epsilon_p(0)$ takes the same value for all $p$. 

If we use the relations (\ref{dotQ1}) and (\ref{dIU}), together with the condition $I=0$, we obtain \cite{EPL.94.60004.2011.Anghel} $\dot Q=\dot U$ and
\begin{equation}
   \kappa \equiv \frac{\partial\dot U_1}{\partial T} + \frac{\partial\dot U_1}{\partial\mu}\left(\frac{d\mu}{dT}\right)_{I=0} . \label{dotQexpr_gen}
\end{equation}
We can observe now already that in general the expression (\ref{dotQexpr_gen}) for the heat conductivity is similar to the expression for the heat capacity, if we replace $\dot U_1$ by $U$, i.e. the energy current by the internal energy. 

Let us now set $\cN=1$ and focus only on one channel conduction. In such a case, $I_{1}(\mu,T)\equiv I_{p=1,1}(\mu,T)$ and we observe that if we set $\epsilon_{p=1}(0)\equiv 0$, then $I_{1}(\mu,T)$ and $\dot U_{1}(\mu,T)$ have the same expressions as the particle number, $N(\mu,T)$, and internal energy, $U(\mu,T)$, respectively, of a FES system of parameter $\alpha$ and constant DOS, $\sigma=\zeta_{p=1}/h$, in equilibrium at temperature $T$ and chemical potential $\mu$. Therefore we can apply here directly the results from the previous section.

Let us drop the subscript 1 from $I_{1}(\mu,T)$ and $\dot U_{1}(\mu,T)$ and rewrite the equilibrium thermodynamics equations of section \ref{thermodynamics}, using the quantities corresponding to the stationary transport. We define again a quantity $y_0$, which satisfies
\begin{subequations}\label{IUpoly}
\begin{eqnarray}
  &&I(\mu,T) \equiv \frac{\zeta\kB T}{h}\log(1+y_{0}), \label{In} \\
  && (1+y_{0})^{1-\alpha}/y_{0} = \rme^{-\beta\mu},  \label{yin} 
\end{eqnarray}
\begin{equation}
  \exp{[(\mu - \alpha h I/\zeta)/\kB T]} = 1- \exp{[-hI/(\zeta\kB T)]},  \label{muI}
\end{equation}
in analogy to eqs. (\ref{N})-(\ref{mu})

From the eqs. (\ref{IUpoly}) and observing that $\dot U(\mu,T)$ has an expression similar to that of $U(\mu,T)$, we obtain 
\begin{eqnarray}
\dot U &=& -\frac{1-\alpha}{2}\frac{hI^2}{\zeta} - (\kB T)^2 \frac{\zeta}{h} Li_2(-y_{0}) = \frac{\alpha hI^2}{2\zeta} + \frac{\zeta(\kB T)^2}{h}Li_2\left(\frac{y_{0}}{y_{0}+1}\right) , \label{dotUechiv2} 
\end{eqnarray}
\end{subequations}
where we used again the Landen's relations.

From Eqs. (\ref{IUpoly}) and using 
\begin{subequations}\label{dy0dmudT}
\begin{equation} 
  \left(\frac{\partial y_0}{\partial T}\right)_{\mu} = -\frac{\mu y_0(1+y_0)}{\kB T^2(1+\alpha y_0)} \label{dy0dT}
\end{equation}
and
\begin{equation} 
  \left(\frac{\partial y_0}{\partial\mu}\right)_{T} = \frac{y_0(1+y_0)}{\kB T(1+\alpha y_0)}\label{dy0dmu}
\end{equation}
\end{subequations}
we calculate
\begin{subequations} \label{dIdTdmu}
\begin{eqnarray}
  \frac{\partial I}{\partial T} &=& 
\frac{\zeta\kB}{h}\log(1+y_{0}) - \frac{\zeta\mu}{hT}\frac{y_{0}}{1+\alpha y_{0}}, \label{dIdT} \\
  \frac{\partial I}{\partial \mu} &=& 
  = \frac{\zeta y_{0}}{h(1+\alpha y_{0})}. \label{dIdmu} 
\end{eqnarray}
\end{subequations}
\begin{subequations} \label{ddotUdTdmu}
\begin{eqnarray}
\frac{\partial \dot U}{\partial T} &=& -(1-\alpha)\frac{hI}{\zeta}\frac{\partial I}{\partial T} - 2k_{\rm B}^2 T \frac{\zeta}{h} Li_2(-y_{0}) + \frac{\mu I(1+y_{0})}{T(1+\alpha y_{0})}, \label{ddotUdT}
\\ 
\frac{\partial \dot U}{\partial\mu} &=&-(1-\alpha)\frac{hI}{\zeta}\frac{\partial I}{\partial\mu} - 2k_{\rm B}^2 T \frac{\zeta}{h} Li_2(-y_{0}) + \frac{I(1+y_{0})}{1+\alpha y_{0}}, \label{ddotUdmu}
\end{eqnarray}
\end{subequations}
%

With eqs. (\ref{dIdTdmu}) and (\ref{ddotUdTdmu}) and using the identity 
\begin{equation}
  \left.\frac{d\mu}{dT}\right|_{I=0} = -\left.\frac{\partial I}{\partial T}\right|_{\mu}\left(\left.\frac{\partial I}{\partial \mu}\right|_{T}\right)^{-1} \label{dmudT_I}
\end{equation}
we can calculate the heat conductivity. Plugging (\ref{dmudT_I}) into (\ref{dotQexpr_gen}) we obtain 
\begin{equation}
  \kappa=\left.\frac{d\dot U}{dT}\right|_{I} = \frac{\partial\dot U}{\partial T}-\frac{\partial\dot U}{\partial \mu}\left.\frac{\partial I}{\partial T}\right|_{\mu}\left(\left.\frac{\partial I}{\partial \mu}\right|_{T}\right)^{-1} \label{kappa_gen}
\end{equation}
which, if we replace $I$ by $N$ and $\dot U$ by $U$, becomes identical to the expression for the heat capacity, $C_V$ (\ref{cv}). Therefore we can transcribe Eq. (\ref{cv}) into an equation for the heat conductivity as 
\begin{equation}
  \kappa = - \frac{I^2h}{T\zeta}\frac{1+y_{0}}{y_{0}}-\frac{2k_{\rm B}^2T\zeta}{h}Li_2(-y_{0}) , \label{kappa_univ} 
\end{equation}
which is independent of $\alpha$ at \textit{any temperature}. In the low temperature limit Eq. (\ref{kappa_univ}) becomes the universal asymptotic expression, $\kappa_0$, which is equivalent to (\ref{cv_limT0}) from equilibrium thermodynamics, but with $\sigma\equiv\zeta/h$. 


The heat flux is related to the entropy flux by the relation $d\dot U_{1,I=0}=Td\dot S_1$, from where we obtain an entropy flux 
\begin{subequations} \label{S_dot_S_defs}
\begin{equation}
  \dot S_1 = \int_0^T \frac{\kappa}{T'}\rmd T', \label{dot_S1_def}
\end{equation}
which is consistent with the definition used in equilibrium thermodynamics, 
\begin{equation}
  S = \int_0^T\frac{C_V}{T'}\rmd T'. \label{S_def_CV}
\end{equation}
\end{subequations}
But since $S$ of (\ref{S_def_CV}) is consistent with the Boltzmann definition (\ref{entropy_conf}) and therefore with the integral of eq. (\ref{SWtot_int}), it follows that the definition (\ref{dot_S1_def}) is also consistent with 
\begin{eqnarray}
  \dot S &=& \frac{\kB \zeta}{h}\int_{0}^{\infty}\big\{[1+(1-\alpha)n(\epsilon)]\log[1+(1-\alpha)n(\epsilon)]\nonumber \\ 
  && -n(\epsilon)\log[n(\epsilon)]-[1-\alpha n(\epsilon)]\log[1-\alpha n(\epsilon)]\big\}d\epsilon \nonumber 
  \\
  &=& -\frac{k_{\rm B}^2 T\zeta}{h} [2Li_2(-y_{0}) + \log{(1+y_{0})}\log{y_{0}}] \label{entropia_flux} ,
\end{eqnarray} 
which represents the flux of the number of configurations of particle populations, $\{n(\epsilon)\}$, from one reservoir to the other. But we notice again that eq. (\ref{entropia_flux}), like eq. (\ref{SWtot_int}), is independent of $\alpha$ at \textit{any temperature}. Therefore the entropy flux is independent of $\alpha$ at any $T$.

\subsection{Microscopic interpretation} \label{transp_micro}

From the r.h.s. of eq. (\ref{dotUechiv2}) we observe that the temperature dependent part of the energy flux $\dot U$ is the excitation energy flux, $\dot U_{B}\equiv\frac{\zeta(\kB T)^2}{h}Li_2\left(\frac{y_{0}}{y_{0}+1}\right)$, which is independent of $\alpha$. For $\alpha=0$, $\dot U\equiv\dot U_{B}$. The microscopic interpretation of this statistics independence is similar to the one given in section \ref{thermodynamics}. 

\begin{figure}[t]
\begin{center}
  \includegraphics[width=10cm]{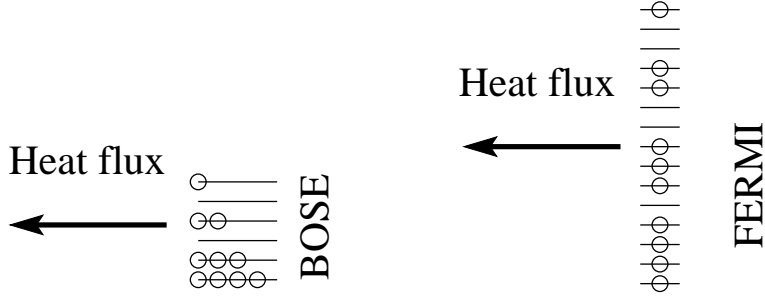}
\end{center}
\caption{The microscopic analysis of the statistics independence of the heat conductivity: there is a one-to-one correspondence between micro-configurations of particles of different statistics which have the same excitation energy, $\dot U_{B,n,1}$, and therefore carry the same heat fluxes.}
\label{excitations_flux}
\end{figure}

As in Ref. \cite{JPA35.7255.2002.Anghel}, one can realize a one-to-one correspondence between micro-configurations of particles that transport the same excitation energies, although they have different exclusion statistics. An example of two micro-configurations like this, one of bosons and one of fermions, is given in figure \ref{excitations_flux}. Since two such micro-configurations transport the same excitation energies and the same particle fluxes, then they transport the same heat fluxes. Moreover, since the entropy flux is the flux of the number of micro-configurations, the entropy flux should also be independent of the statistics \cite{EPL.94.60004.2011.Anghel}.

\section{Conclusions} \label{conclusions}

I reviewed the thermodynamic equivalence of systems of the same, constant density of states (DOS) and the independence of statistics of the heat conductivity of one-dimensional (1D) transport channels. I showed that these two classes of apparently different physical phenomena have many similarities and practically the formulae that describe the equilibrium thermodynamics of constant DOS systems can be directly transcribed for the transport of 1D channels. 

I also presented the microscopic explanation of these universal features of FES gases. I showed that the thermodynamic equivalence is caused by the independence of statistics of the particle excitation spectra, whereas the universality of the heat conductivity of 1D channels is due to the independence of statistics of particle excitation fluxes. 

In conclusion, the similarity between the mathematical formulation of the equilibrium thermodynamics of constant DOS systems and the mathematical formulation of the transport properties of 1D channels is a reflection of the microscopic similarities between these two types of physical phenomena.

\ack

The work was supported by the Romanian National Authority for Scientific Research, CNCS-UEFISCDI project PN-II-ID-PCE-2011-3-0960 and project PN09370102/2009. The travel support from the Romania-JINR Dubna collaboration project Titeica-Markov and project N4063 are gratefully acknowledged.

\section*{References}


\end{document}